# Session-based Communication for Vital Machine-to-Machine Applications

Marc-Olivier Arsenault[1], Hanen Garcia Gamardo[1],
Kim-Khoa Nguyen[2,] Mohamed Cheriet[2]

[1]Ericsson Canada inc., Town of Mount-Royal, Canada
`{marc-olivier.arsenault, hanen.garciagamardo}@Ericsson.com`
[2]École de Techologie Supérieure, Montréal, Canada
`{knguyen}@synchromedia.ca`
`{mohamed.cheriet}@etsmtl.ca`

**Abstract.** Although the machine to machine (M2M) communication has been emerging in recent years, many vendors' specific proprietary solutions are not suitable for vital M2M applications. While the main focus of those solutions is management and provisioning of machines, real-time monitoring and communication control are also required to handle a variety of access technologies, like WiFi and LTE, and unleash machine deployment. In this paper, we present a new architecture addressing these issues by leveraging the IP Multimedia Subsystem (IMS) deployed in operator's networks for RCS and VoLTE.

## 1 Introduction

The Machine to Machine (M2M) communications have been around for years. It used to be a very specific scenario where vendors offer very specific proprietary solutions for specific verticals. However, with several standards in place and business advantage provided by interworking solutions, M2M environment is becoming a more horizontal solution for any industry to connect their devices [9].

One of the main features offered by new generation of networks, like 5G, will be the presence of the M2M communications. The realm of M2M communications presents itself with incredible business opportunities [13] across all industries, and for engineers building solutions to connect 50 billion devices this will be a commensurable challenge.

The behavior of a machine's communication is very different regarding human communication. Human communication is often predictable in terms of time (day, night) and volume (frequency) [9]. Thus, operators can encode human voice with predefined codecs using specific bandwidth. They can then calibrate their networks to receive X number of calls per second with Y concurrent calls. Unfortunately, there is no such thing in the M2M environment where some machines may connect once a week to send only few bytes, while some other may maintain a constantly open connection to stream videos for example. In [13], the authors claim driverless car will use 50GB of bandwidth per hour to be driven remotely. All this devices will need connectivity to send their data, and mobile carriers can offer mobility to the devices anywhere there is coverage. Also, their networks are more reliable than the public best-effort Internet.

There are many solutions for the M2M communications. Some over the top (OTT) solutions [19] offer communication services that connect devices to the Internet to route data, and provide other services like security, broadcast, store and forward, database storage, etc. Regarding the enormous number of devices (e.g. up to 50 billion in 2020 [21]), and many of them are not IP-based, additional communication methods are required to complement Internet-based solutions.

In the Telco world, some larger scale scenarios, like OneM2M [17] standard, have been proposed to afford such requirements. OneM2M architecture is based on a push-and-pull mechanism where devices can push data to a central database and authorized services can retrieve data when they need it. Many existing M2M solutions use this push-and-pull model [3] [4] [17]. So far, there is no stream-oriented M2M mechanism which is required for critical and vital applications such as security monitoring, health and life monitoring, real-time recognition, etc. that is controllable, monitored and secured.

In addition, other M2M network capabilities are also required to fulfill application needs. First, an agnostic network is required to receive connections from different underlying access technologies, such as wired or wireless access. The M2M network architecture must also be able to accept devices and recognize them wherever they are connecting from. Second, M2M network should be provided with a scalable addressing scheme which is auto configured when devices are deployed on the field. Such scheme must be simple and allows the connection of billions of devices.

Some previously proposed solutions like [6] [12] offered IP Multimedia Subsystem (IMS) [1] oriented connectivity, which reuse IMS sessions to connect devices together. These solutions deal partially with the general M2M communication problem by creating a session-based communication but do not really address the two aforementioned concerns.

This paper will present a Machine-to-Machine Session Based Communication (MSBC) architecture that provides Telco-grade stream-oriented communications for critical applications and offers security, reliability, network agnostic and scalable addressing scheme.

The next section describes the MSBC architecture and discusses how M2M communication could benefit from this architecture. The third section presents an experimental environment used to build and test the MSBC prototype. Finally, we conclude the paper and present future work.

## 2  MSBC Architecture

The proposed MSBC solution leverages the existing operator's network with a minimal impact on infrastructure and network topology. It is an IMS-based service that provides session based M2M communications by reusing technologies already in place for RCS and VoLTE [1] [7] [8], and proven standards protocols for session management like SIP and MSRP [20] [10]. The IMS allows the MSBC to create session based communication between machines and servers. It provides lots of controls and monitoring while ensuring a certain security in machine authentication.

Based on the latest IMS specification [1], MSBC can be deployed and tested in any standard environment without any major impact. One of the major benefits of using the IMS is the network agnosticity as the IMS is designed to interconnect subscribers from different access technologies or even from multiple accesses at the same time.

**2.1 The MSBC Architecture:**

The proposed MSBC solution is a multi-tier architecture, composed of three main components, as shown in Figure 1: the local gateway (LGW) on the local domain, the M2M Interconnect Server (M2M-IS) in the operator domain and the application service gateway (ASGW) in the M2M services provider domain.

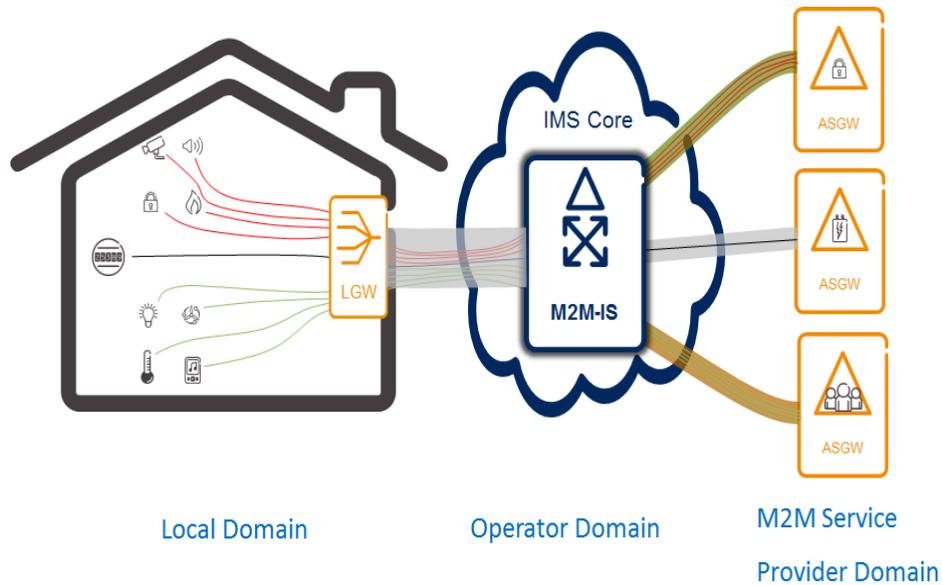

**Fig.1**. Machine-to-Machine Session Based Communication overview

The local gateway (LGW) provides the connectivity to the machine domain (so called local domain). In the IMS world, the LGW represents the User Equipment (UE) through a SIP subscriber. The LGW is connected to the IMS and manages network connectivity and switches over different networks without the need of synchronization with any application server.

The LGW has two functions. The first one is to handle the communication with different devices within the local domain. The second function is to manage the communications between the local domain of devices and the M2M network in the operator domain. All the sensors, motors, detector and any other type of connected devices will be connected to the LGW which will handle the data communication of the network. The local domain connections can be different for all the devices like Bluetooth, ZigBee, etc. and there is no limit of connectivity as long as the hardware and software in the gateway has the technology to handle them. The LGW also handles services on top of the connectivity. It is responsible for maintaining connection, security establishment, and mapping addresses. When the LGW receives the data from the devices, it uses the second function of the gateway to send data to network.

The service application gateway (ASGW) plays a similar role as the LGW, however it only establishes communication with a single application and not with many different devices like the LGW. As well, the ASGW is responsible to authorize the communication of different devices in the network. The ASGW is implemented in the service provider's application server that handles communication with specific machines.

The third component is the M2M Interconnect Server (M2M-IS). This server manages connection between the CT attached to the LGW and all related ASGW, so data from devices can reach the corresponding application server.

Most of M2M communications technologies are based on device's unique identifier (ID) in order to be able to get addressed by other services in a network, like MSISDN on GSM, or IP address on the Internet [2]. In the MSBC network, we define Connected Thing Identifier (CTID) as a serial number or any agreed identity mechanism, which is basically the only identifier recognized in the MSBC network and used not only for getting addressed as destination but for addressing their counterpart on the network as the source of information.

Routing in this network is done via a logical link, called wire. The wires are virtual connections between one device and one application server, creating a unique path that allows data to flow from one end to the other.

With this wire concept, the network becomes a source-based network, where identified devices are assigned to a wire, enabling them to communicate with the appropriate application server. There is no need for the device or the gateways to know the destination address, the CTID is enough for the M2M-IS to route data to the right application server.

Such mechanism removes coupling between the device and the application server, thus it suppresses the need of network addressing configurations on the devices which facilitates massive deployments of machines and adds flexibility to service providers regarding the management of the services offered to the machines, like migrating from one service to another without changing device configuration.

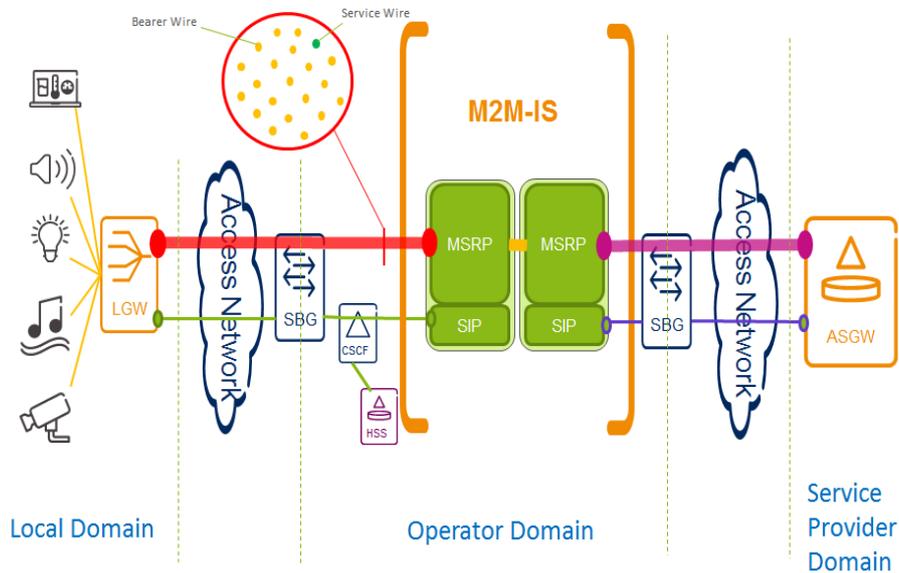

**Fig. 2.** MSBC architecture and session establishment

Figure 2 shows an established session between the LGW in the local domain and the M2M-IS. The session establishment is the same process for the ASGW in the service provider domain and the M2M-IS, which simplifies the setup of such sessions.

Within a session, there are active bearer wires and the service wire. The active bearer wires represent connected devices and the service wire is used to setup the communication between the gateways and the M2M-IS, which includes: authentication, commissioning, decommissioning of CT, monitoring and other control features to ensure the session is up at all times.

Once the LGW detects a new device on its local network, it will issue a bearer wire request for that device using the wire service. In turn, the M2M-IS will assign a new wire after the authorization of the device is cleared by the ASGW with the application server. From this point, all data sent by the device will be packaged in a wire packet with a wire identifier and sent to the M2M-IS which forwards the packet to the ASGW, and finally delivers to the application server.

The MSBC architecture reuses the IMS core with an add-on of an extra service, called the M2M Interconnect Server. The M2M-IS is designed to be deployed in the service layer of an IMS network. In this way it limits the impacts and facilitates the integration on operator's network, and at the same time, it inherits the network agnostic characteristic of the IMS service layer. In other words, the deployment process of an infrastructure like MSBC, is to find an operator with a configured IMS architecture, add the M2M-IS node, and create subscribers in the operator's HSS with the address of M2M-IS configured by the service trigger.

When all these steps are done, a service provider can simply contact the operator to register for subscribers to this system and starts using the MSBC.

To create an LGW or ASGW session, which can be considered as IMS UE, a SIP INVITE will be initiated to the IMS core. Both gateways are registered in the IMS

HSS as a subscriber with a specific M2M subscription profile. This M2M profile triggers the M2M-IS that acts as an AS proxy. When the M2M-IS receives a SIP INVITE, it will complete the SIP dialog by sending a 200 OK status to establish a direct session with the gateway. During this SIP dialog the M2M-IS and the gateway will negotiate an MSRP session using the SDP like defined in RCS. The MSRP session will then be established between the gateway and the M2M-IS. Figure 3 presents the session establishment.

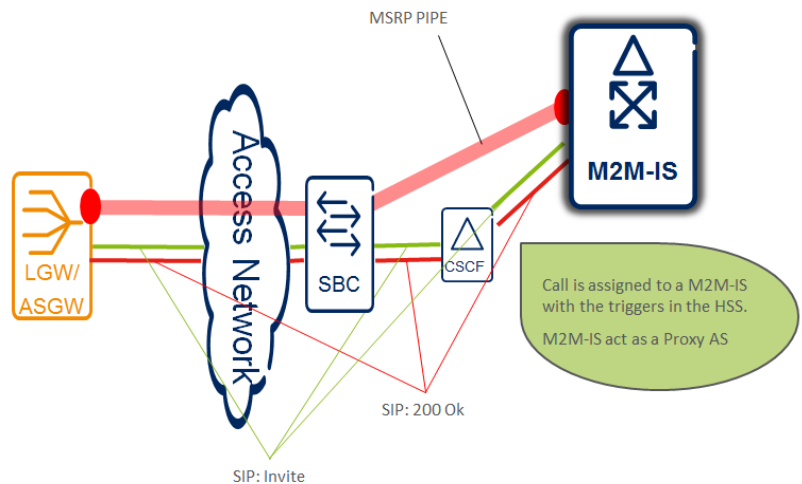

**Fig. 3.** MSBC session establishment

MSRP [10] is a very good candidate for M2M stream oriented payload protocol. It is an already existing and accepted standard in the telecom world. It is a de facto payload protocol for messaging in IMS and 4G networks. Nodes in the networks already support the protocol, and real world issues such NAT traversal is already resolved by using nodes like BGF and SBC [1]. MSRP also provides technical benefits; it is a reliable protocol and can be secured when using MSRP over TLS, which is the case in our network. It also offers a reporting mechanism ensuring all the messages are received properly.

MSRP is a text-based protocol, like SIP or SDP. All nodes in a session flow can act like B2BUA (Back to Back User Agent) and change some part of the messages or the session to ensure privacy when it goes into another network, like in roaming situations.

### 2.2 Benefits:

Today, network agnosticity has become a real requirement for M2M communication. There is a need for machine to move from a network to another based on different parameters. Considering a home automation system which might use

Internet connection for regular behavior, unfortunately when connection is lost or a security breach is detected, the system needs to switch to radio access (like LTE), reconnect to the MSBC, and sends alert. This sample scenario shows why such network agnostic requirement emerges. Even when connected from a different access, IP, the MSBC recognizes the device and forward it to the appropriate application server. The proposed MSBC offered on top of IMS, provides a solution that meets this requirement. As long as you can reach the IMS core, you are able to use the MSBC features. Based on network access technology different behavior can be applied. In case of radio access, communication is encrypted by the underlying network, thus there is no need to encrypt MSRP with TLS. However, in case of local cable access, such precaution needs to be taken. On the other hand, local cable access offers almost free bandwidth compared to radio access, thus more data can be sent, including bandwidth-greedy unimportant updates that were held in standby.

Regarding these features, the MSBC would be a strategic asset for operators, and even if it could be deployed as an Over the Top service, it will rely on the underlying operator's network to provide QoS avoiding suffer from best effort Internet access. Together with security requirements, the implicit operator's security measures in their access network optimize resource consumption of the explicit TLS fallback mechanism while performing on a public IP connection. Therefore, the MSBC is a more suited solution for an operator's already enabled IMS network.

In the MSBC architecture, we introduced a communication method where a speaker is identified and paired with a listener by the network, based on the observer pattern. The speaker does not decide who should be listening to him or to whom he has to speak to. It is the network pairing mechanism that makes this decision.

Device manufacturer and service provider do not need to hardcode or discover each other. A device simply needs to know its own CTID and the MSBC will find the appropriate AS. The network will do mapping between these two. It also prevents a device of setting the destination address, which increases the security. The only control they have is on their own identity. This addressing mechanism allows managing billions of devices in a secure way.

The introduction of the MSRP for the user plane allows the MSBC to offer a session based communication between machines. This is significantly different regarding existing solutions based on push-and-pull (PaP) mechanisms like the well-established SMS [11] or the first proposed architecture from OneM2M [17].

Since the proposed MSBC architecture is IMS-based, SIP is required as signaling protocol to establish sessions. Indeed, IMS uses SIP to find devices and to communicate with them. Like MSRP, using SIP brings some benefits. Being a well-known and widely used protocol, it can be implemented in operator's network with no major changes. All VoLTE networks have already been configured to handle SIP signaling on all sections of network, from access point to core network. SIP allows routing in complex topology with the Route/Via system. It is a text-based protocol that can be modified to keep sender's privacy, etc.

## 3 Experimental results

To test and prove the proposed MSBC architecture, a prototype has been built, as depicted in Figure 4.

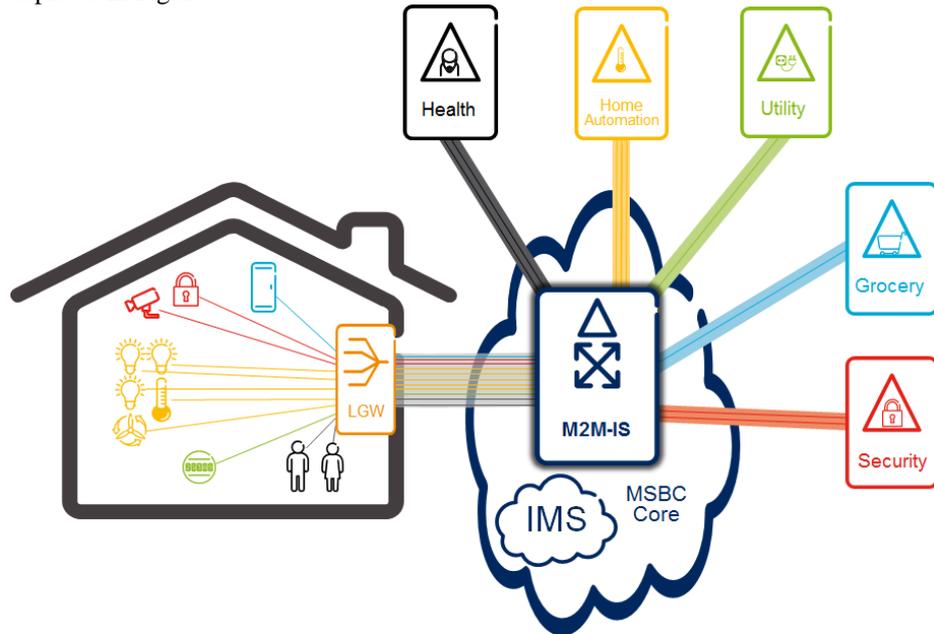

**Fig. 4.** MSBC network prototype

**3.1 The MSBC Core**

The MSBC core includes two components, the IMS core and the M2M-IS as presented in section 2. The IMS core is a Standard Ericsson IMS solution [5]. The core topology is hidden by an Ericsson SBC, which acts as a B2BUA on the SIP dialog and as media pass through for MSRP session. SIP encryption/decryption is handled by the SBC, but MSRPS encryption/decryption passes through the SBC and is handled by the M2M-IS.

The M2M-IS was developed in JAVA and runs on a virtual machine using Glassfish [18] as JAVA virtual machine (JVM). The M2M-IS was developed to be scalable and to accept millions of devices, thus all the component in the architecture of this node were designed using the micro-services design pattern for cloud applications [16]. All SIP INVITE that come from a device are handled by a small instance of the M2M-IS. Computing resources can easily be added when the M2M-IS needs to be scaled up. The M2M-IS uses a modified version of the Ericsson MSRP stack optimized to handle TLS/SSL encryption and an open source SIP stack [15].

### 3.2 The MSBC Gateways:

The LGW and ASGW were implemented in Java using the well-known JSIP project [15] which publishes the Reference Implementation for JAIN-SIP-1.2 as the SIP stack, and a proprietary Ericsson MSRP stack.

The LGW and ASGW are proposed as a library that could be embedded by developers of M2M solutions. The library offers a very simple API to connect devices to the MSBC network. The LGW library offers the following services as static method: Open(receiver) - Open a connection to the MSBC network, and assign the data receiver, Transmit(ctid, data) - Transmit data of a device identified by its CTID and Close() - Close connection to the MSBC network. It also offers a callback interface (data receiver) to process data received for the device identified by its CTID

The ASGW library offers the same API plus another call in the callback interface for CT authorization. These two libraries are used in the simulation environment to create and use the MSBC communication channel.

### 3.3 Simulation environment

In order to test the MSBC prototype, we have built a test environment to simulate the implementation of a real-life use case, under the form of a smart home use-case as depicted in Figure 4. The smart home simulation is a Play Mobil [14] dollhouse where sensors, motors, actuators and other things are added, as shown in Figure 6. It emulates what a real house could look like. We created five service providers; each offers a specific service to the smart house. Each service comes with a set of devices installed in the house. All communication between the devices and the service providers is done via the MSBC's network.

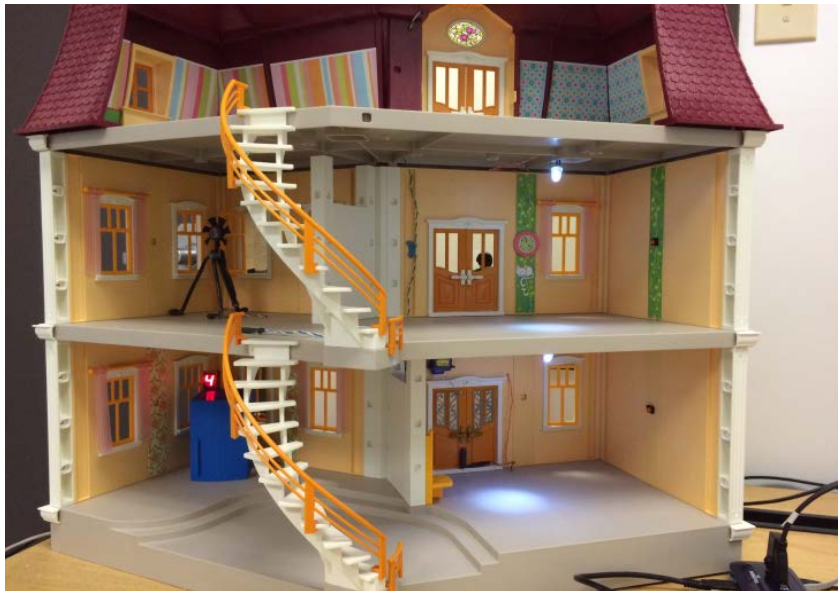

**Fig. 6.** Play Mobil house with sensors

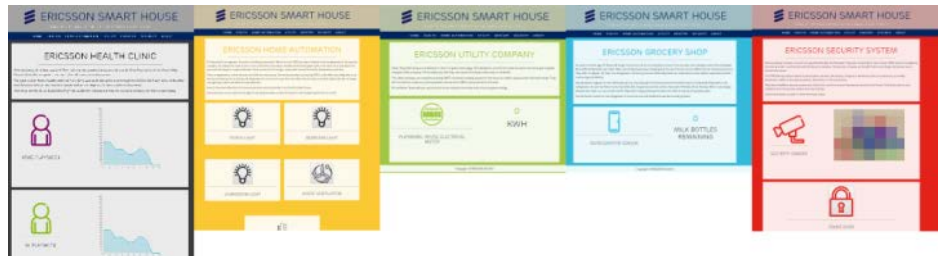

**Fig.7.** Service Provider Web Portal

The web GUI interfaces of these five smart home services are shown in Figure 7. From left to right:
1. Health care (Black): Represents a health clinic where a doctor could monitor client heartbeat. Associated sensors are small heart monitors on the Play Mobil characters
2. Home Automation (Yellow): Represents a home automation system where a home owner remotely controls his home equipment such as: light, ventilation etc. Associated sensors are three lights, one fan and one thermometer.
3. Utility (Green): Represents a utility company meter system to measure electricity consumption and to do the billing. The associated sensor is a small electric meter.
4. Grocery Store: (Blue): Represents a grocery service that monitors the refrigerator content to automatically order goods when low on reserve. The sensors represent a counter of milk bottle in the refrigerator (shown in Figure 6 in blue with the 4 LED segment)
5. Security (Red): Represents a security monitoring service that controls security system including doors control, alarms and cameras. The associated sensors are the door system to open or close, and a camera.

### 3.4 Test scenario

We used the system presented above to test the MSBC. Following tests were run to validate functionality of the proposed architecture.

| Test | Detail | Result |
|---|---|---|
| Add/Remove CT (easy deployment) | Added and remove virtual wire to the MSBC to test the wire establishment process | < 75 ms to establish and remove a wire |
| Transmission of data on existing wire | Once a wire is established, we run traffic on this wire | Transfer time < 50ms to send a wire and receive confirmation that AS received it |
| Clean shutdown | Test the disconnection of the house with a clean shutdown | All wires are cleanly disconnected, M2M-IS instances is shut down properly and Service provider is warned as expected. No ghost process or artifact remaining |
| Lost connectivity - Watchdog | Test connection loss without clean shutdown by unplugging the interface | As expected, the watchdog detects the link failure and starts the shutdown procedure |
| Switch between LTE and Wi-Fi | Test network agnosticity of the MSBC. We did all the connections using a local Internet and a 3G Dongle | All connection worked exactly the same from both connection points. The service provider could not detect from where the clients were connected and the addresses were exactly the same. |
| Swap from one AS to another (easy migration service provider) | Test a Service provider server switch by disconnecting the ASGW and reconnecting it from a different IP address | No packet were lost, the MSBC buffered the packet and sent them back as expected |

**Table 1** Prototype tests result

Through the testing, we observed that the MSBC prototype is addressing the needs for M2M communication. Furthermore, network agnosticity and addressing capabilities will provide a scalable, secure and efficient solution. Next step could be to address the network switch at run time.

## 4 Conclusion

We have presented a session based communication for vital machine–to-machine applications, as an alternative to transaction-based or push-and-pull established M2M solutions. The proposed MSBC offers a solution for the vital M2M communication which requires a real-time monitoring while offering a solution for device addressing and network agnosticity. Implemented in the IMS network service layer, the proposed solution enables flexibility, scalability and portability in machine deployment.

The first version of MSBC has been designed for a traditional deployment of IMS services, but today's technologies evolution challenges this implementation of services as a shortsighted view, and opens the question of what native cloud deployment services are optimal for MSBC. This evolution has been introduced on the MSBC roadmap and will be part of our future work.

# References

[1] 3gpp. "Ip Multimedia Subsystem (Ims)." 3gpp, 2015. 13.3.0 vols. Print.
[2] 3gpp. "Numbering, Addressing and Identification." 13.2.02015. Print.
[3] Blum, N., et al. "Application-Driven Quality of Service for M2m Communications." Intelligence in Next Generation Networks (ICIN), 2011 15th International Conference on. 2011. Print.
[4] Elmangoush, A., et al. "Design Aspects for a Reference M2m Communication Platform for Smart Cities." Innovations in Information Technology (IIT), 2013 9th International Conference on. 2013. Print.
[5] Ericsson. "Ericsson Ims."  2015. Web.
[6] Foschini, L., et al. "M2m-Based Metropolitan Platform for Ims-Enabled Road Traffic Management in Iot." Communications Magazine, IEEE 49.11 (2011): 50-57. Print.
[7] GSMA. "Rich Communication Suit." 2015. Vol. 5.3. Print.
[8] GSMA. "Voice and Video Calls over Lte."  2015. Web.
[9] Holler, J., et al. From Machine-to-Machine to the Internet of Things: Introduction to a New Age of Intelligence. Elsevier Science, 2014. Print.
[10] Ietf. "Rfc4975 - the Message Session Relay Protocol (Msrp)."  (2007). Print.
[11] Ietf. "Uri Scheme for Global System for Mobile Communications (Gsm) Short Message Service (Sms)." rfc57242010. Print.
[12] Jaewoo, Kim, et al. "M2m Service Platforms: Survey, Issues, and Enabling Technologies." Communications Surveys & Tutorials, IEEE 16.1 (2014): 61-76. Print.
[13] McCourt. Travis C, Leopold. Simon. Internet of Things, a Study in Hype, Reality, Distruption, Growth. USA: Raymond James, 2014. Print.
[14] Mobil, Play. "Large Grand Mansion."  2015. Web.
[15] Mudumbai Ranganathan, phelim, and vralev. "Jainsip - 1.2."  2010. Web.
[16] Newman, Sam. Building Microservices. " O'Reilly Media, Inc.", 2015. Print.
[17] oneM2M. "Standards for M2m and the Internet of Things." ATIS.oneM2M.TS0002V101-2015. OneM2M2015. Print.
[18] Oracle. "Slassfish Server - World's First Java Ee 7 Application Server."  (2015). Print.
[19] Pubnub. "Build Realtime Apps & Take Websockets to the Next Level | Pubnub."  2015. Web.
[20] Rosenberg, J., et al. Rfc 3261: Sip: Session Initiation Protocol2002. Print.
[21] Vestberg, Hans. Ceo to Shareholders: 50 Billion Connections 2020 Print.